\documentstyle[preprint,prl,aps,psfig]{revtex}
\tightenlines
\begin{document}
\draft
\title{Far infrared absorption in triangular and square quantum dots:
characterization of corner and side modes}
\author{Manuel Val\'{\i}n-Rodr\'{\i}guez, Antonio Puente and
Lloren\c{c} Serra}
\address{Departament de F\'{\i}sica, Universitat de les Illes Balears,
E-07071 Palma de Mallorca, Spain}
%
%
\maketitle
\begin{abstract}
The far-infrared absorption of triangular and square two-dimensional
quantum dots is studied by means of time simulations of the density
oscillations within the time-dependent local-spin-density approximation.
The absorption is spatially analyzed using a
local-response theory that allows the identification of {\em corner}
and {\em side} modes in the geometric nanostructures.
The evolution with a vertical magnetic field
of varying intensity is also discussed.
\end{abstract}
\narrowtext

\section{Introduction}

The quantum-dot field is recently focussing much attention on the
analysis of symmetry unrestricted nanostructures, both from the
experimental and theoretical point of view. An important issue is the
influence of the dot geometry on the edge modes that manifest in the
far-infrared absorption. It is well known that parabolically
confined dots can only absorb energy at the
frequencies of the confining parabolas (generalized Kohn's
theorem \cite{Bre89,Mak90}).
However, either departures from a pure radial quadratic law or the
introduction of an additional angular dependence in the confining
potential may cause the fragmentation of the Kohn modes and
the emergence of a more complicated pattern.

At first, the great majority of theoretical approaches to quantum-dot
absorption were restricted to the circular symmetry case, solving the 
random-phase-approximation (RPA) equations by fully exploiting the 
analytical angular dependence of the single-particle orbitals
\cite{Bro90,Gud91,Ser97}.
Although very successful, this scheme is not easily applicable to
symmetry-unrestricted nanostructures and different approaches 
have been proposed.
In Ref.\ \cite{Mag99} Magn\'usd\'ottir and Gudmundsson address the problem
within the Hartree theory by solving the resulting RPA matrices
in the single-particle basis of the circular problem. The enormous
matrix dimensions make the calculation feasible only for a rather small
electron number. Alternatively, Ullrich and Vignale \cite{Ull00} have 
used a perturbative method to describe the absorption of elliptic dots.
A different approach was used by two of us in Ref.\ \cite{Pue99},  
consisting in the time
simulation of the mean field density oscillations that follow an
appropriate initial perturbation and from which one may extract the
oscillation frequencies. This {\em real time} calculations 
do not rely on any symmetry restriction having also been used to
analyze excitations in deformed atomic nuclei and metal clusters
\cite{Flo78,Yab96}.

In this work we concentrate on the description of the absorption in
triangular and square shaped quantum dots using the time-dependent
local-spin-density approximation (TDLSDA). The extension to low magnetic
fields within this theory is considered as in Ref.\ \cite{Ull00}.
We show that the deformation leads to a departure from the
Kohn modes which is peculiar to the shape. At low magnetic fields
the spectra possess a rich structure being possible to identify the 
presence of peak {\em anticrossings} associated with the deviation 
from circular
shape, similar to those found experimentally by Demel et al. \cite{Dem90}.
To further clarify the origin of the fragmentation, an analysis of the
{\em local} absorption at different points of the nanostructure
is given. This allows to associate each excitation with a particular
absorption pattern, thus giving a more physical interpretation of the peaks. 
In particular, a distinction between {\em corner}
and {\em side} modes can be established in this way. The magnetic field
is shown to introduce an {\em angular average} of the patterns leading,
for increasing intensity, to sizeable shape modifications of the absorption
distribution.

Section II of the paper is devoted to the characterization of the
ground state for the square and triangular quantum dots. In Sec.\ III
we address the description of the far infrared absorption presenting in
different subsections the formalism and global spectrum (III.A), the
local absorption (III.B) and the angular average and magnetic evolution of
the absorption patterns (III.C). The conclusions are given in Sec.\ IV.

\section{System characterization}

We focus our attention on semiconductor nanostructures of similar
sizes and different polygonal shapes, in order to discuss the
differences introduced by the geometry. It is assumed that the electron
positions are restricted to the $xy$ plane where, in addition, an external
potential is confining them to an electronic island. Different shapes are
obtained by using the following potential, proposed by
Magn\'usd\'ottir and Gudmundsson \cite{Mag99},
\begin{equation}
V_{ext}(r,\theta) = \frac{1}{2}\omega_0^2 r^2 (1+\alpha_p \cos(p\theta))\; ,
\label{eq1}
\end{equation}
where we have used planar polar coordinates $(r,\theta)$ and the 
standard effective-atomic-unit system \cite{units}.
The potential
parameter $\omega_0$ gives the parabola curvature while $p$ and $\alpha_p$
determine the deviation from circular symmetry.
In this work we shall restrict to triangular and square shapes. These
are the two simplest geometries that deviate from a pure parabolic 
behaviour and whose absorption is therefore not fixed by the generalized 
Kohn's theorem. 
A common parabola frequency $\omega_0=0.5$ a.u.\ has been chosen
both for the triangle and square while the deformation parameters have
been fixed to $\alpha_3=0.31$ and $\alpha_4=0.15$
respectively, in order to obtain a proper triangle and box shape,
as can be seen in Fig.\ 1.

We have described the electronic structure within density-functional
theory in the LSDA.
This approach has been used to address a variety of nonuniform
electronic systems --- such as atoms, molecules and clusters ---
and quantum dots are not an exception for
many authors have used this scheme to analyze dot
properties \cite{Fer94,Pii98,Kos97,Hir99}.
Obviously, for our present purposes a symmetry unrestricted solver is
required. This can be obtained with different
techniques \cite{Ull00,Kos97,Hir99}
such as diagonalization in a plane-wave basis or perturbative treatements
based on a circular solution. We have resorted to the method developed
by two of us in Ref.\ \cite{Pue99}, based on the uniform discretization
of the $xy$ plane and the associated transformation of the diferential
Kohn-Sham equations to be solved with iterative
techniques. 
This method has also been applied by us to the study
of quantum rings \cite{Val00} and dot molecules \cite{Val01}.

In what follows we have restricted the present study to nanostructures
containing an intermediate number of electrons ($N=12$) for which
LSDA is expected to provide a reasonable description of electron-electron
interactions. Figure 1 shows that the electronic distribution closely
follows the shape of the external potential, clearly of triangular or
square form. 
However, there are also conspicuous density peaks localized
either at the corners or the interior of the structure. They 
are of course attributed to the formation of delocalized electronic
orbitals, and to quantum effects associated with the presence of
boundaries in the potential, similar to the well-known Friedel 
oscillations at a metal surface.

\section{FIR Absorption}

\subsection{Formalism and spectra}

The description of excitations in deformed nanostructures constitutes a
highly nontrivial task because of the lack of symmetry which does not 
allow the analytic integration of angular variables as in circular systems.
In the present work we have used the TDLSDA to obtain the time
evolution, following an initial perturbation, of relevant expectation
values.
The single-particle orbitals $\{ \varphi_i({\bf r})\}$ evolve in
time as
\begin{equation}
i{\partial\over\partial t} \varphi_{i\eta}({\bf r},t) =
h_\eta[\rho,m]\, \varphi_{i\eta}({\bf r},t)\;,
\label{eqh}
\end{equation}
where $\eta=\uparrow,\downarrow$ is the spin index, and total density
and magnetization are given in terms of the spin densities
$\rho_\eta({\bf r})=\sum_i{|\varphi_{i\eta}({\bf r})|^2}$, by
$\rho=\rho_\uparrow+\rho_\downarrow$ and
$m=\rho_\uparrow-\rho_\downarrow$, respectively. The Hamiltonian $h_\eta$
in equation (\ref{eqh}) contains, besides of kinetic energy,
the confining and
$v^{\em (conf)}({\bf r})$, Hartree 
$v^{(H)}({\bf r})=\int{d{\bf r}' \rho({\bf r}')/{|{\bf r}-{\bf r}'|}}$
exchange-correlation
$v^{(xc)}_\eta({\bf r})=
{\partial\over\partial\rho_\eta}{\cal E}_{xc}(\rho,m)$ potentials.
The exchange-correlation energy density ${\cal E}_{xc}(\rho,m)$ has been
described as in Refs.\ \cite{Ull00,Pue99}. The Zeeman energy 
$\pm (1 / 2) g^*\mu_B B$ 
\cite{Zeem} is also included in Eq.\ \ref{eqh}.

An initial perturbation of the ground state orbitals
$\varphi'({\bf r})={\cal P}\varphi({\bf r})$ models the interaction with the
external field. For dipole excitations, corresponding to FIR absorption
in nanostructures,
the unitary operator ${\cal P}$ is represented by a rigid translation
in an arbitrary direction $\hat{\bf e}$, specifically 
\begin{equation}
{\cal P} = \exp{\left[
i {\lambda\, \hat{\bf e}\cdot{\bf p} } \right] }\; ,
\label{eq16}
\end{equation}
where the $\lambda$ parameter has been taken small enough in order to
keep the system response in the linear regime.
After the initial excitation we keep track of
$\langle{\bf r}\rangle(t)$
from where a frequency analysis provides the
absorption energies and their associated intensities \cite{Pue99},
characterized by the strength function
$S_{\hat{\bf e}}(\omega)$,
\begin{eqnarray}
{\cal D}_{\hat{\bf e}}(\omega) &=&
\hat{\bf e}\cdot\int{dt\, e^{i\omega t}\, \langle{\bf r}\rangle(t)}\; ,
\nonumber \\
S_{\hat{\bf e}}(\omega)&=&\left|{\cal D}_{\hat{\bf e}}(\omega)\right|
\; .
\end{eqnarray}
Usually, one is interested in an average over different oscillation
directions, corresponding to the absorption of nonpolarized radiation.
This implies to average Eq.\ (4) over all $\hat{\bf e}$ directions. 
Taking
into account that in the small amplitude limit the strength in a particular
direction only depends on the initial amplitude given in that direction
the averaged strength may be obtained as \cite{epl}
\begin{eqnarray}
S_{av}(\omega) &\equiv&
\frac{1}{2\pi} \int_0^{2\pi}{d\theta\,
S_{\hat{\bf e}}(\omega)}\nonumber\\
&=& \frac{1}{2\pi}
\int_0^{2\pi}{d\theta\, 
\left\vert \cos(\theta)^2 {\cal D}_{x}(\omega) +                        
           \sin(\theta)^2 {\cal D}_{y}(\omega) \right\vert }\; .
\end{eqnarray}
It can easily be shown that  
for geometries with two or more equivalent symmetry axes 
$S_{\hat{\bf e}}(\omega)$ actually does not depend on ${\hat{\bf e}}$
and therefore it coincides with $S_{av}(\omega)$.

Figure 2 displays the averaged spectrum obtained at $B=0$ for the square
and triangle. Notice that both cases show 
a two peak spectrum within LSDA, although the interpeak separation
is much higher for the triangle than for the square.
This can be attributed to the higher level separation in the
triangle. In fact, in the triangular case the single-particle levels 
exhibit a shell-like structure with gaps between groups
of close levels, while
for the square level straggling is somewhat higher. This reflects in the
distribution of particle-hole transitions also shown in Fig.\ 2.

The evolution of the absorption spectra with the intensity of the static
magnetic field is shown in Fig.\ 3. In general, the magnetic field
induces a splitting of the $B=0$ peaks, as is well known for the
circular parabola \cite{Mak90} where the two frequencies are
$\omega_\pm=\sqrt{\omega_0^2+\omega_c^2/4}\pm\omega_c/2$. 
For the square and triangular dots the situation  
is more involved, since the $B$-induced
splitting goes along with geometry effects. Figure 3
indicates that there is a repulsion amongst the branches that originate from
the $B=0$ modes, showing a zone of avoided crossings or {\em anticrossings}
of the collective peaks. These anticrossing areas are very sensitive to the
geometry, showing complicated patterns with more than just two peaks.
Indeed, for the square it roughly covers the interval $0.3-0.7$ T while
for the triangle this region is more extended lying between $0.5-2$ T.
A similar difference is found when comparing
the magnetic field $B_{pol}$ at which the ground state spin polarizes 
from $S=0$ to $S=1$. The square first transition lies at $B_{pol}=0.5$~T
while for the triangle it is $B_{pol}=2.7$~T.
We also notice that the minimum energy gap between the two main  
branches is much smaller for the square than for the triangle.

\subsection{The local absorption}

To further analyze the origin of the fragmented absorption in squares and
triangles we have computed the {\em local} absorption in a way similar
to that used in Ref.\ \cite{Mol00} in the context of
excitonic states in semiconductors. This technique allows
us to study the global absorption spectrum taking into account the different
contributions from each spatial point, depending on how the electronic density
is oscillating in time in the vicinity of that point. In practice, we 
simulate this position-dependent probe by introducing
a local weighting function (a narrow Gaussian centered on each point) and
look for the oscillation frequencies of the convoluted density as a 
function of the probe position.
After an initial dipole perturbation each
local signal will evolve differently in time, thus manifesting the differences
in absorption on a local scale.  Figure 4 shows the local absorption for the
two modes at $B=0$ of the square and triangle when the system is excited 
along the $x$-axis ($\hat{\bf e}=\hat{\bf x}$). 
As a reference we also show the circular parabolic case, namely, $\alpha_p=0$ 
in Eq.\ (\ref{eq1}).

The density in a circular parabola oscillates rigidly at the parabola frequency
(Kohn's theorem) and from Fig.\ 4 we notice that the absorption indeed
localizes on the edges along the oscillation direction ($x$-axis), as was
to be expected. Turning now to the square geometry we notice that
the dominanting mode at $\omega=0.49$, very close to the parabola frequency $\omega_0=0.5$,
spatially localizes along the edges in the polarization direction in a way quite
similar to the circular case. On the contrary, the lower square mode at
$\omega=0.42$ very clearly deviates from this situation showing four 
maxima at the corners of the nanostructure.
This result suggests a classification of the two excitations as
{\em side} and {\em corner} modes, respectively. This classification
is confirmed in the case of the triangle, where we see that the $\omega=0.73$
mode basically absorbs on one of the triangle sides while the $\omega=0.44$
absorbs on the opposite corner. The distinction between side and corner
modes will be further commented in the next subsection where we shall present
the angular average of the absorption patterns.

A remarkable property seen in Fig.\ 4 regards the spatial self-organization
of the absorption modes. Indeed, we realize that both in the square and
the triangle the absorption maxima
for a given mode lie at regions where the absorption for the other mode
is strongly quenched. We may
interpret this as a {\em mode competition} for an efficient distribution
over the available nanostructure surface.

\subsection{Magnetic field dependence}

The magnetic field introduces a Coriolis-like force that makes
the oscillation axis rotate in time. As a consequence, the initial chosen
direction becomes unimportant since the magnetic field will automatically
change it in time. We can take advantage of this effect to obtain the
averaged absorption patterns of Fig.\ 4 over polarization directions by
switching on a very small magnetic field. Figure 5 shows the angularly 
averaged patterns corresponding to the same modes displayed in Fig.\ 4. 
The distinction between corner and side
modes is very clear in the case of the triangle: for $\omega=0.44$
the mode is of corner type and for $\omega=0.73$ it is of side character.
For the square this distinction is not as clear because the corners always
absorb more than the side centers. Nevertheles, the corner character is
somehow higher for the low energy mode $\omega=0.42$ than for the  high 
$\omega=0.49$ one; in
fact the former shows vanishing absorption at the side centers while the
latter has a much smoother angular dependence.

When the magnetic field is increased additional fragmentation, with a
complicated distribution of the absorption, appear both in the square
and the triangle (cf.\ Fig. 3). Because of the very large number of possible
cases, we shall only select the more illustrative 
examples in the anticrossing areas.
Our aim is to discern the absorption pattern evolution with applied magnetic
field and in particular, to check whether there is a relation between the
energy repulsion of modes and their spatial localization. Figures 6 and 7
show the evolution with $B$ of the pattern for the two branches with the
higher absorption, that repel each other, in the square and triangular dot, 
respectively.
We see in Fig.\ 6 that indeed for the square a clear localization
takes place at $B=0.4$~T (the center of the anticrossing region) 
at the corners for low energy branch and at the sides for the 
high energy one.
In the triangle, a more conspicuous 
side and corner character of the branches is retained for all 
magnetic fields due, as discussed above, to the existence of a greater 
energy gap between the different modes.
Figure 7 shows the patterns for $B=1$~T as well as the corresponding
energy in the spectral range.
The three lower patterns give rise, when added according to their relative 
strength, to the global strength distribution displayed 
in the rightmost panel which exhibits a clear corner identity. Conversely,
the upper branch possess a markedly side character. 
The same same trends have been obtained 
in the full range of magnetic fields corresponding to Fig.\ 3 for the triangle.

\section{Conclusions}

The FIR absorption of a quantum dot with a triangular or
square shape has been analyzed taking into account the effect of an 
applied static magnetic field, by means of real-time simulations in a
spatial mesh within the LSDA formalism.
The fragmentation of the absorption due to the
noncircular shape has been emphasized. In particular, the existence of
a rich absorption distribution for intermediate
magnetic fields, associated with the
repulsion between modes in the so-called anticrossing energy interval
has been analyzed.

The absorption pattern corresponding to the several excitation modes, 
for a given magnetic field, has been obtained from the local density 
oscillations. These
patterns allow us to easily interpret the physical character of each
mode. In particular, we have identified the existence of corner and side
modes in the nanostructure. 
In the anticrossing intervals the absorption patterns of different
modes do not significantly overlap, a result that we atribute to a
mode competition for the available dot surface as well as a
mode spatial repulsion, consistent with the energy repulsion leading to
the anticrossing effect.

\acknowledgements
This work was supported by the spanish DGESeIC,
Grant No.\ PB98-0124.


\begin{figure}[h]
\caption{Density distributions for the square and triangular dots
with 12 electrons considered in this work (see Sec.\ II).
Upper plots display the densities as three
dimensional surfaces while lower plots show the corresponding contour
lines. }
\end{figure}

\begin{figure}[f]
\caption{Absorption spectra for the square and triangle. Solid line
corresponds to the TDLSDA while the dashed line shows the non-interacting
response. For clarity the non-interacting response has been slightly 
shifted in the vertical direction. 
The right plots show the relative position of the lowest energy 
levels in each case with the arrow indicating the position of the 
highest-occuppied orbital.}
\end{figure}

\begin{figure}[f]
\caption{Magnetic field evolution of the absorption spectrum of the
square and triangle shown in a grey colour scale, with black and white
indicating high and low aborption levels, respectively.}
\end{figure}

\begin{figure}[f]
\caption{Local absorption patterns in the circle, square and triangle 
for an oscillation along the $x$-axis and the energies indicated
in each panel.}
\end{figure}

\begin{figure}[f]
\caption{Average over polarization directions of the same modes of Fig.\
4 obtained by switching on an small vertical magnetic field.}
\end{figure}

\begin{figure}[f]
\caption{Magnetic field evolution of selected absorption patterns
within the anticrossing region of the square dot. The magnetic field
and energy allow to exactly identify
what mode of Fig.\ 3 is being considered in each case. }
\end{figure}

\begin{figure}[f]
\caption{Spatial absorption patterns corresponding to different modes
of the triangular nanostructure for $B=1$~T. The arrows join each spatial
pattern to its corresponding excitation energy in the spectral interval. 
See text for more details.}
\end{figure}



\begin{references}

\bibitem{Bre89}
L. Brey, N. F. Johnson, and B. I. Halperin,
Phys.\ Rev.\ B {\bf 40}, 10647 (1989).

\bibitem{Mak90}
P. A. Maksym and T. Chakraborty,
Phys.\ Rev.\ Lett.\ {\bf 65}, 108 (1990).

\bibitem{Bro90}
D. A. Broido, K. Kempa, and P. Bakshi,
Phys.\ Rev.\ B {\bf 42}, 11400 (1990).

\bibitem{Gud91}
V. Gudmundsson and R. R. Gerhardts,
Phys.\ Rev.\ B {\bf 43}, 12098 (1991);
V. Gudmundsson and J. J. Palacios,
Phys.\ Rev.\ B {\bf 52}, 11266 (1995).

\bibitem{Ser97} Ll.\ Serra and E. Lipparini,
Europhys.\ Lett.\ {\bf 40}, 667 (1997);
Ll.\ Serra {\em et al.}, Phys.\ Rev.\ B {\bf 59}, 15290 (1999).

\bibitem{Mag99} I. Magn\'usd\'ottir and V. Gudmundsson,
Phys.\ Rev.\ B {\bf 60}, 16591 (1999).

\bibitem{Ull00} C. A. Ullrich and G. Vignale,
Phys.\ Rev.\ B {\bf 61}, 2729 (2000).

\bibitem{Pue99} A. Puente and Ll.\ Serra,
Phys.\ Rev.\ Lett.\ {\bf 83}, 3266 (1999).

\bibitem{Flo78} H. Flocard, S. E. Koonin and M. S. Weiss,
Phys.\ Rev.\ C {\bf 17}, 1682 (1978).

\bibitem{Yab96} K. Yabana and G. F. Bertsch,
Phys.\ Rev.\ B {\bf 54}, 4484 (1996); L. Mornas, F. Calvayrac, E. Suraud and 
P.-G. Reinhard, Z.\ Phys.\ D {\bf 38}, 73 (1996).

\bibitem{Dem90} T. Demel, {\em et al.},
Phys.\ Rev.\ Lett.\ {\bf 64}, 788 (1990).

\bibitem{units}
We use $\hbar=m=e^2/\kappa=1$, which for the GaAs effective mas 
$(m=0.067\, m_e)$ and dielectric constant $\kappa=12.4$ imply 
an energy and length units of $\approx 12$ meV and 
$\approx 97$~\AA, respectively. 

\bibitem{Fer94} M. Ferconi and G. Vignale,
Phys.\ Rev.\ B {\bf 50}, 14722 (1994).

\bibitem{Pii98} M. Pi, M. Barranco, A. Emperador, E. Lipparini and 
Ll. Serra, Phys.\ Rev.\ B {\bf 57}, 14783 (1998).

\bibitem{Kos97} M. Koskinen, M. Manninen, S. M. Reimann,
Phys.\ Rev.\ Lett.\ {\bf 79}, 1389 (1999).

\bibitem{Hir99} K. Hirose and N. S. Wingreen,
Phys.\ Rev.\ B {\bf 59}, 4604 (1999).

\bibitem{Val00} M. Val\'{\i}n-Rodr\'{\i}guez, A. Puente, and Ll.\ Serra,
Eur.\ Phys.\ J D {\bf 12}, 493 (2000).

\bibitem{Val01} M. Val\'{\i}n-Rodr\'{\i}guez, A. Puente, and Ll.\ Serra,
Proceedings of the ISSPIC10 conference, Eur.\ Phys.\ J. D (2001), 
to be published.

\bibitem{Zeem} $\mu_B$ denotes the Bohr magneton and $g^*$ the effective 
gyromagnetic factor which in bulk GaAs takes a value of $-0.44$.

\bibitem{epl} A. Puente and Ll.\ Serra,
Europhys.\ Lett., in press.

\bibitem{Mol00} C. D. Simserides {\em et al.},
Phys.\ Rev.\ B {\bf 62}, 13657 (2000);
O. Mauritz {\em et al.},
Phys.\ Rev.\ B {\bf 62}, 8204 (2000).

\end{references}
\end{document}